\documentclass[oneside]{amsart}
\usepackage[T1]{fontenc}
\usepackage[latin1]{inputenc}

\makeatletter

\providecommand{\LyX}{L\kern-.1667em\lower.25em\hbox{Y}\kern-.125emX\@}

 \theoremstyle{plain}
 \newtheorem{thm}{Theorem}[section]
 \numberwithin{equation}{section} 
 \numberwithin{figure}{section} 
 \theoremstyle{plain}
 \newtheorem{cor}[thm]{Corollary} 
 \theoremstyle{plain}
 \newtheorem{lem}[thm]{Lemma} 
 \theoremstyle{plain}
 \newtheorem{prop}[thm]{Proposition} 
 \theoremstyle{definition}
 \newtheorem{defn}[thm]{Definition}
 \theoremstyle{definition}
  \newtheorem{example}[thm]{Example}
 \theoremstyle{remark}
 \newtheorem*{rem*}{Remark}
\def\be{\begin{equation}\label}
\def\ee{\end{equation}}

\usepackage[T1]{fontenc}
\usepackage[latin1]{inputenc}

\makeatletter

\usepackage[T1]{fontenc}
\usepackage[latin1]{inputenc}

\makeatletter

\usepackage[T1]{fontenc}
\usepackage[latin1]{inputenc}

\makeatletter

\usepackage[T1]{fontenc}
\usepackage[latin1]{inputenc}

\makeatletter

\usepackage[T1]{fontenc}
\usepackage[latin1]{inputenc}

\makeatother
\makeatother
\makeatother
\makeatother

\begin{document}

\footnotetext{AMSC 1991: Primary 46E39, Secondary 30C60; Keywords: Sobolev spaces,
Embedding theorems, Quasiconformal mappings, rough domains }

\author{Vladimir Gol'dshtein}

\address{Department of Mathematics\\
 Ben-Gurion University of the Negev\\
 P.O.B. 653, Beer-Sheva 84105, Israel. \\
 E-mail: vladimir@bgumail.bgu.ac.il}

\author{Alexander G. Ramm}

\address{Department of Mathematics\\
 Kansas State University\\
 Manhattan, KS 66506-2602, USA.\\
 Email: ramm@math.ksu.edu}

\title{Compactness of the embedding operators for rough domains.}

\maketitle

\begin{abstract}
 New classes of non-smooth bounded domains $D$, for which
the embedding operator from $H^1(D)$ into $L^2(D)$
is compact, are introduced. These classes include in particular
the domains whose boundary locally are graphs of $C-$functions,
but also contain much larger classes of domains.
Examples of non-smooth domains for which the above embedding is
compact are given. Applications to scattering by rough
obstacles are mentioned.
\end{abstract}

\section {Introduction}

In this paper we prove some results about
compactness
of embedding operator \( H^{1}(\Omega )\to L^{2}(\Omega )
\) for rough bounded domains, that is, for domains with non-smooth
boundaries which do not satisfy the usual for the embedding
theorems conditions, such as cone condition, Lipschitz domains,
and extension domains (Ext-domains). First, we prove compactness of
the embedding operators for ``elementary'' domains which can be
approximated by Lipschitz domains in the sense described
below (see the paragraph above Lemma 1.2). 
This class \( ET \)
of ``elementary'' domains is larger then
 the  known classes of domains used in embedding theorems.
Let us give some bibliographical discussion.
In \cite{Ra} a necessary and sufficient condition for compactness
of the embedding operator is given in an abstract setting.
A version of this result is  presented in the Appendix.
 Compactness of the embedding
operator  for bounded domains with ``segment property''
is proved in  \cite{A}. In  \cite{F}
it was shown that the class of  domains with
``segment property''  coincides with the class \( C  \) of domains
whose boundaries are locally graphs of continuous functions.
Compactness of the embeddings for the class \( C  \)
was proved  in \cite{CG}. The reader can found an interesting discussion
of  these results in \cite{PM}. The class \( ET \) is much larger than
the class \( C  \) and includes (in the two-dimensional case) bounded
domains  whose boundaries are generated locally by the graphs of
piecewise-continuous functions with ``jump''-type discontinuity
at a finite number of points. Boundaries of the 
domains of class \( ET \)
can have singularities more complicated than the
``jump''-type singularities 
(see example 
\ref{jump}).

Using a Sobolev-type lemma for the union of ``elementary'' domains of the
class \( ET \) we extend this result to domains of the class \( T \)
which are finite unions of the ``elementary'' domains. Simple examples
demonstrate that boundary of a bounded domain of class  \( T \)
can have countably many connected components (see example \ref{component}). This is impossible for
the classes of domains used in embedding theorems earlier (compare, for
example, 
classes \( C  \) and \( E \) in  \cite{F},  \cite{PM},
\cite{E} with our class
\( T \)).

Our construction can be generalized. First, we construct a class
of ``elementary'' domains with the compactness property for the
embedding operator. Secondly, we extend this compactness property
to finite unions of ``elementary'' domains. This scheme is used for
quasiisometrical (the class \( L \)) and \( 2- \)quasiconformal
  (the class \( Q \))
cases. Our class \( L \) includes the Fraenkel 
class \( E  \). Let us explain this. Note that \( E  \)
in [9] is not extension domains.
According to [9],p.411,  any domain \( \Omega  \) of
 class \( E \) is locally \( C^{1} \)
-diffeomorphic at any boundary point to a domain of  class \(
C \) and \( \partial \Omega =\partial \overline{\Omega } \)
where \( \overline{\Omega } \) 
is the closure of \( \Omega  \). Any domain
of our class \( L \) is a 
finite union of domains that locally quasiisometrically
equivalent at any boundary point to domains of class \( ET
\) . The condition
\( \partial \Omega =\partial \overline{\Omega } \) is not 
necessary for the domains in this
class. For example, if $\Omega$ is a disc with an
extracted radius, then $\Omega$ is a domain of the
class \( L \), but  \( \partial \Omega \neq \partial
\overline{\Omega } \). 

Our class \( Q \) is much larger then the class \( L \) and 
includes domains
with some ``anisotropic'' behavior of 
their boundaries (see section 4.3 for
detailed explanation).

Our results allow one to use the results in \cite{R} and \cite{RR} on the
existence and uniqueness of the solutions to the
scattering problem in the exterior of rough obstacles and consider larger 
class of rough obstacles in scattering theory than it was done earlier.

\section {Abstract result}
In this section we prove some results
which give conditions  for the  compactness of
an embedding operator, 
and use these results in a study of compactness
of the embeddings of Sobolev spaces. 
An abstract 
necessary 
and sufficient condition for the  compactness of
an embedding operator is 
 proved in \cite{Ra}.
Let $H_1$ and $H_2$ be Hilbert spaces and $H_1$ $\subset H_2$ . 
Here the
embeddings are understood as the
 set-theoretical inclusions and the inequalities $||u||_1\ge
||u||_2$ are assumed, 
where $||u||_j:=||u||_{H_j}$. 
Suppose 
that \( T_{s} \) , \( s\in (0,1) \) is a family of closed
subspaces
of \( H_{2} \) 
and 
 \( T_{\sigma}\subset T_{s } \) for \( s<\sigma  \).

In our applications $T_s=L^2(D_s),$ 
where $D_s\subset D$, $D_s\subset D_{\sigma}$ for $s<\sigma.$
We assume below (see lemma 1.2) that
a domain $D,$ for which we study the compactness of the
embedding operator from $H^1(D)$ into $L^2(D),$ contains a Lipschitz
subdomain $G_s,$  $D_s\subset G_s\subset D$.
Let $P_s$ be the orthogonal projection onto $T_s$ in $H_2$,
$i:H_1 \to H_2$ be the embedding operator, and $i_s:=P_si$.
Let us state two results.  The above assumptions and notations
are not repeated. The first result is obvious.

\begin{prop}
If the operator \( i:H_{1}\rightarrow H_{2} \) is compact,
 then
the operator
\( i_{s}\) is compact for any \( s\in (0,1) \)
.
\end{prop}

\begin{proof}  
 If $i$: $H_1\to H_2$ is compact, 
then $i_s$
is a composition of a bounded linear operator $P_s$ and a
compact operator
$i$, so $i_s$ is compact.
\end{proof}

The following proposition is  used in the proof of
proposition 1.4 below.

\begin{prop} \label{abs}
 If the 
following conditions hold:

$1)\text{ }i_s\text{ is compact for all }s\in (0,1) \text{and}
$

$
2)\,||u||_2\le a(s)||u||_1+b||P_su||_{2}, \,\,
a(s)>0,\,\lim_{s\rightarrow 0}a(s)=0,
\,\,  b>0\text{ for any }u\in H_1,
\label{1}$
then the embedding $i: H_1 \to H_2$ is compact.
\end{prop}

\begin{proof} 

Choose a sequence \( \{s_{m}\} \) such that \( \alpha (s_{m})<\frac{1}{m} \).
Denote by \( P_{m} \) the projection \( P_{s_{m}} \). If \( i_{s} \) is compact
then \( \parallel u_{n}\parallel _{1}=1 \) implies \( \parallel P_{s}(u_{n})\parallel _{2}\leq 1 \)
and for any \( m \) there exists a subsequence \( u_{n,m} \) and a number
\( n(m) \) such that \( \parallel P_{m}(u_{n,m}-u_{n_{1},m})\parallel _{2}<\frac{1}{m} \)
for any \( n,n_{1}\geq n(m) \). Without loss of generality we can suppose that
the sequence \( u_{n,m_{1}} \) is a subsequence of \( u_{n,m} \) and \( n(m)<n(m_{1}) \) for \( m<m_{1} \) . Therefore
$$
 \parallel P_{m}(u_{n(m),m}-u_{n(m_{1}),m})\parallel _{2}<\frac{1}{m} 
$$
for any \( m \) and for any \( m_{1}>m \) .

For the subsequence \( u_{m}:=u_{n(m),m} \) condition 2) implies \( \parallel u_{m}-u_{m_{1}}\parallel _{2}\leq \alpha (s_{m})\parallel u_{m}-u_{m_{1}}\parallel _{1}+b\parallel P_{m}(u_{m}-u_{m_{1}})\parallel _{2} \).
By choice of the subsequence \( \{u_{m}\} \) this implies \( \parallel u_{m}-u_{m_{1}}\parallel _{2}<(1+b)\frac{1}{m} \)
for any \( m \) . We proved convergence of \( \{u_{m}\} \) in \( H_{2} \).
Because the original sequence \( \parallel u_{n}\parallel _{1}=1 \) was arbitrary
compactness of the operator \( i \) proved.

\end{proof}

We apply this abstract result to Sobolev spaces. Suppose that \(
D\subset R^{n} \)
is a bounded domain and \( \{D_{s}\} \), \( 0<s<1 \), is a family
of  subdomains
such that 
\( D_{s}\subset D_{\sigma} \)
for any \( s <  \sigma \) 
and for any \( s \) there exists a Lipschitz
domain
\( G_{s} \) such that \( {D_{s}}\subset G_{s}\subset D \). A
bounded
domain
is a Lipschitz domain if its boundary is locally graph of 
a Lipschitz function. 

\begin{lem}
\label{2trace}Suppose that \( \{u_{n}\} \) is 
a bounded sequence in \( H^{1}(D) \).
Then there exists a subsequence \( \{u_{n_{k}}\} \)
of the sequence \( \{u_{n}\} \) which converges in \( L^{2}(D_{s}) \),
i.e.  $i_{s}: H^{1}(D_{s}) \to L^{2}(D_{s})$ is compact for all
$s\in (0,1)$.
\end{lem}
\begin{proof}
One takes the Lipschitz domain \( G_{s} \)  such that \(
D_{s}\subset
G_{s}\subset D \).
By the known embedding theorem for Lipschitz domains  the embedding
\( H^{1}(G_{s})\to L^{2}(G_{s}) \) is compact. 
Since \( D_{s}\subset G_{s}\subset D \),
one obtains the conclusion of the lemma.
\end{proof}

\medskip

\begin{prop} \label{pabs} 
If the following condition hold
$$ 
||u||_{L^{2}(D)} \le a(s)||u||_{H^{1}(D)}+
b||u||_{L^{2}(D_{s})},\quad
a(s)>0,\,\lim_{s\rightarrow 0}a(s)=0,\quad b>0,
$$
for any $u\in H^{1}(D)$.

Then the operator $i:$ $H^{1}(D) \to L^{2}(D)$ is
compact

\end{prop}
By  Lemma 2.3 $i_{s}: H^{1}(D_{s}) \to L^{2}(D_{s})$ is compact for all
$s\in (0,1)$. Hence the claim follows from Proposition 2.2.

\medskip

In section 3-4 we describe classes of domains for which the
conditions of
the proposition 2.4 are satisfied.

\section{Domains of class \( T \).}

Below we denote a domain by $\Omega$.
The main purpose of this section is to prove 
compactness of the embedding operators
\( H^{1}(\Omega )\to L^{2}(\Omega ) \) for domains of the class \( T \) 
which we describe below. Domains of the class  \( T \) are finite unions
of elementary domains of the class  \( ET \) whose
boundaries are locally graphs
of ``good'' functions: 
these domains can be approximated by Lipschitz
subdomains in such a way, that conditions of Proposition \ref{pabs} hold. For
example, in the two-dimensional case the function is
``good'' if
it is piecewise-continuous with discontinuity points of ``finite jump''
type.

In the first part of this section we describe exactly 
classes  \( T \) and  \( ET \).
In the second part we derive an auxiliary one-dimensional 
inequality. This inequality
is not new, but its proof is. It is a version of Agmon inequality \cite{A}
adopted 
for our purposes.
In the final part of this section we prove 
compactness of the embedding operator 
for domains of class \( T \)  using the results of
section 2.

\subsection{Preliminaries.}

Let \( x\in R^{n} \), \( x=(x_{1},x_{2},...,x_{n}) \) and  \( Q_{n}=[0,1]^{n} \) 
be the standard closed cube in \( R^{n} \). Denote \( x^{'}:=(x_{1},x_{2},...,x_{n-1}) \).

A bounded
function \( f:Q_{n-1}\to R \) is an admissible function if $f$ is continuous on a set
\( C\{f\} \subset  Q_{n-1} \) such that 
\( \mu (Q_{n-1}\setminus C(f))=0 \), where \( \mu \) is \( n-1 \)-dimensional 
Lebesgue measure.
Denote by  \( Int {A} \)  the set of all
interior points of \( A. \).

\begin{defn}
We call \( U:=Int\{Q_{n}+(0,...,0,f(x'))\} \) a standard elementary domain
if the function \( f \) is admissible.
\end{defn}
Let \( Q_{n,h}=[h,1-h]^{n} \) be a subcube of the standard cube \( Q_{n} \).
If \( U:=Int\{Q_{n}+(0,...,0,f(x'))\} \) is an elementary domain then denote \\
\( U_{h}=Int\{Q_{n,h}+(0,...,0,f(x'))\} \). 

\begin{defn}
We  call a standard elementary domain \( U \) a standard elementary domain
of class \( ET \) if for any \( 0<h<1/3 \) there exists a Lipschitz domain
\( V_{h} \) such that \( U_{h}\subset V_{h}\subset U \). We  call \( U \)
an elementary domain of class \( ET \) if it is an image of a standard
elementary domain of  class \( ET \) under a affine invertible mapping of
\( R^{n} \) onto \( R^{n} \).
\end{defn}
\begin{example}
Suppose that \( f:[0,1]\to R \) is a piecewise continuous bounded function
with finite number of discontinuity points \( x_{1},x_{2},..,x_{k} \) and at
any discontinuity point the function \( f \) has right and left limits (i.e.
any discontinuity points are ``jump'' points). The domain \( U:=Int\{Q_{2}+(0,f(x))\} \)
is a standard elementary domain of the class \( ET \). 
\end{example}
\begin{proof}
It is obvious that \( U \) is a standard elementary domain. Fix \( 0<h<1/3 \).
The open set \( W_{h}=Int(U\setminus U_{h}) \) is a finite union of domains
\( U_{i}=(x_{i-1},x_{i})\times (1-h+f(x),1+f(x)) \) and 
domains \( V_{i}=(x_{i-1},x_{i})\times (f(x),h+f(x)) \),
\( i=1,...,k+1 \), \( x_{0}=a \), \( x_{k+1}=b \). 
Join any two points \( (x_{i-1},1-h/2+\lim _{x\rightarrow x^{+}_{i-1}}f(x) \),
\( (x_{i},1-h/2+\lim _{x\rightarrow x_{i}^{-}}f(x) \) by a smooth curve \( \alpha _{i} \)
and any pair \( (x_{i-1},h/2+\lim _{x\rightarrow x^{+}_{i-1}}f(x) \), \( (x_{i},h/2+\lim _{x\rightarrow x_{i}^{-}}f(x) \)
by a smooth curve \( \beta _{i} \) . The set \( \partial U\setminus \partial W_{h}\cup (\cup _{i=1}^{k=1}\alpha _{i})\cup (\cup _{i=1}^{k=1}\beta _{i}) \)
is a closed Lipschitz curve that is the boundary of a Lipschitz domain \( V_{h} \).
By construction \( U_{h}\subset V_{h}\subset U \) . Therefore \( U \) is a
standard elementary domain of the class \( ET \). 
\end{proof}
\begin{example} \label{jump}
Suppose that \( f:[0,1]\rightarrow R \) is a piecewise-continuous bounded
function
with countably many isolated discontinuity points \(
x_{1},x_{2},..,x_{k},... \)
and at any discontinuity point the function \( f \) has right and left limits
(i.e. any discontinuity points are ``jump'' points). Suppose also that the
sequence \( \{x_{k}\} \) converges to \( x_{0} \). The domain \(
U:=Int\{Q_{2}+(0,f(x))\} \)
is a standard elementary domain of class \( ET \). 
\end{example}
\begin{proof}
Because \( f \) is continuous in \( x_{0} \) for any \( 0<h<1/3 \) the open
set \( W_{h}=Int {U\setminus U_{h}} \) is a finite union of domains of the same
type as in example 3.3. Therefore the domain \( U:=Int\{Q_{2}+(0,f(x))\}
\)
is a standard elementary domain of the class \( ET \). 

\end{proof}
 
\subsection{One-dimensional inequality}

\begin{lem}
\label{2one} If \( u\in H^{1}((-h,h)) \), then

\[
|\Vert u\Vert _{L^{2}((0,h))}-\Vert u\Vert _{L^{2}((-h,0))}|\leq \sqrt{2}h\Vert \frac{du}{dt}\Vert _{L^{2}((-h,h))}.\]

\end{lem}
\begin{proof}
Since smooth functions are dense in \( H^{1}((-h,h)) \) it is sufficient to
prove the desired estimate only for smooth functions \( u\in H^{1}((-h,h)) \).
Integrating the inequality \( |u(t+h)-u(t)|^{2}\leq (\int _{t}^{t+h}|\frac{du}{ds}(s)|ds)^{2}\leq (\int _{-h}^{h}|\frac{du}{ds}(s)|ds)^{2} \)
with respect to \( t \) over the segment \( [-h,0] \) and using the H\"{o}lder
inequality we obtain
\[
\int _{-h}^{0}|u(t+h)-u(t)|^{2}dt\leq h(\int _{-h}^{h}|\frac{du}{dt}(t)|dt)^{2}\leq 2h^{2}\int _{-h}^{h}|\frac{du}{dt}(t)|^{2}dt.\]
 For any normed space \( X \) and any \( x,y\in X \) the following inequality
holds
\[
|\Vert x\Vert -\Vert y\Vert |\leq \Vert x-y\Vert .\]
 Combining this inequality with previous one we obtain
\[
|(\int _{-h}^{0}|u(t+h)|^{2}dt)^{1/2}-(\int _{-h}^{0}|u(t)|^{2}dt)^{1/2}|\leq (\int _{-h}^{0}|u(t+h)-u(t)|^{2}dt)^{1/2}\leq \]

\[
\sqrt{2}h(\int _{-h}^{h}|\frac{du}{dt}(t)|^{2}dt)^{1/2}.\]

Because \( \int _{-h}^{0}|u(t+h)|dt=\int _{0}^{h}|u(t)|dt \) we have finally
\[
|\Vert u\Vert _{L^{2}((0,h))}-\Vert u\Vert _{L^{2}((-h,0))}|\leq \sqrt{2}h\Vert \frac{du}{dt}\Vert _{L^{2}((-h,h))}.\]

\end{proof}
\begin{cor}
If \( u\in H^{1}((-h,h)) \), then
\[
\int _{0}^{h}|u(t)|^{2}dt\leq 2\int _{-h}^{0}|u(t)|^{2}dt+4h^{2}\int _{-h}^{h}|\frac{du}{dt}(t)|^{2}dt,\]
 and
\[
\int _{-h}^{0}|u(t)|^{2}dt\leq 
2\int _{0}^{h}|u(t)|^{2}dt+4h^{2}\int _{-h}^{h}|\frac{du}{dt}(t)|^{2}dt.\]

\end{cor}
\begin{proof}
Using Lemma \ref{2one}, one gets:
\[
\int _{0}^{h}|u(t)|^{2}dt\leq 
[\int _{-h}^{0}|u(t)|^{2}dt)^{1/2}+\sqrt{2}h(\int
_{-h}^{h}|\frac{du}{dt}(t)|^{2}dt)^{1/2}]^{2}\leq \]

\[
2\int _{-h}^{0}|u(t)|^{2}dt+4h^{2}\int _{-h}^{h}|\frac{du}{dt}(t)|^{2}dt.\]

\end{proof}
\begin{prop}
\label{2lineq} If \( u\in H^{1}((a,b)) \), then
\[
\int _{a}^{b}|u(t)|^{2}dt\leq 3\int _{a+h}^{b-h}|u(t)|^{2}dt+4h^{2}\int _{a}^{b}|\frac{du(t)}{dt}|^{2}dt.\]
 for any \( h<\frac{b-a}{4} \).
\end{prop}
\begin{proof}
By the previous corollary
\[
\int _{a}^{b}|u(t)|^{2}dt\leq \int _{b-h}^{b}|u(t)|^{2}dt+\int _{a+h}^{b-h}|u(t)|^{2}+\int _{a}^{a+h}|u(t)|^{2}dt\leq \]

\[
2\int _{b-2h}^{b-h}|u(t)|^{2}dt+\int _{a+h}^{b-h}|u(t)|^{2}+2\int _{a+h}^{a+2h}|u(t)|^{2}dt+\]

\[
4h^{2}\int _{a}^{a+h}|\frac{du}{dt}(t)|^{2}dt+4h^{2}\int _{b-h}^{b}|\frac{du}{dt}(t)|^{2}dt\leq \]

\[
3\int _{a+h}^{b-h}|u(t)|^{2}dt+4h^{2}\int _{a}^{b}|\frac{du}{dt}(t)|^{2}dt.\]

\end{proof}

\subsection{Compactness for elementary domains of class \( ET \) }

\begin{prop}
\label{ecomp}If \( U=(a,b)\times \Gamma _{f} \) is an elementary domain of
the class \( ET \) , then the embedding operator \( i:H^{1}(U)\Rightarrow L^{2}(U) \)
is compact.
\end{prop}
\begin{proof} It is sufficient to prove this proposition for a standard
elementary
domain of class \( ET \).

Fix \( h<\frac{1}{3} \) and choose a sequence \( \{u_{n}\}\subset H^{1}(U) \),
\( \Vert u_{n}\Vert _{H^{1}(U)}\leq 1 \) for all \( n \). Since \( H^{1}(U) \)
is a Hilbert space, one may assume without loss of generality that the sequence
\( \{u_{n}\} \) weakly converges in \( H^{1}(U) \) to some function \( u_{0}\in H^{1}(U) \).

Using Proposition \ref{2lineq} for almost all \( x' \) in the domain of
definition
\( Q_{n-1} \) of continuous function \( f \) we get
\[
\int _{f(x')}^{1+f(x')}|u_{n}(x',t)-u_{0}(x',t)|^{2}dt\leq 3\int _{f(x')+h}^{f(x')+1-h}|u_{n}(x',t)-u_{0}(x',t)|^{2}dt\]

\[
+4h^{2}\int ^{1+f(x')}_{f(x')}|\frac{d(u_{n}-u_{0})}{dt}(x',t)|^{2}dt.\]

Integrating this inequality over \( Q_{n-1} \) we obtain
\[
\int _{U}|u_{n}(x)-u_{0}(x)|^{2}dx\leq 3\int _{U_{h}}|u_{n}(x)-u_{0}(x)|^{2}dx\]

\[
+4h^{2}\int _{U}|\nabla (u_{n}-u_{0})(x)|^{2}dx.\]

Therefore all conditions of  Proposition  \ref{pabs} hold and the embedding operator
\( i \) is compact. 

\end{proof}

\subsection{Compactness for domains of class \protect\protect\protect\protect\protect\( T \protect \protect \protect \protect \protect \).}

\begin{defn}
\emph{A domain \( \Omega  \) belongs to class \( T \) if it is a finite union
of elementary domains of class \( ET \).}

First, we prove a Sobolev type lemma for compact embeddings.
\end{defn}
\begin{lem}
\label{sob}Let \( \Omega _{1} \) and \( \Omega _{2} \) be such domains that
embedding operators \( H^{1}(\Omega _{1})\to L^{2}(\Omega _{1}) \)
and \( H^{1}(\Omega _{2})\to L^{2}(\Omega _{2}) \) are compact, then
the embedding operator \( H^{1}(\Omega _{1}\cup \Omega _{2})
\to L^{2}(\Omega _{1}\cup \Omega _{2}) \)
is also compact.
\end{lem}
\begin{proof}
Choose a sequence \( \{w_{n}\}\subset H^{1}(\Omega _{1}\cup \Omega _{2}) \),
\( \Vert w_{n}\Vert _{H^{1}(\Omega _{1}\cup \Omega _{2})}\leq 1 \) for all
\( n \). Let \( u_{n}:=w_{n}|\Omega _{1} \) and \( v_{n}:=w_{n}|\Omega _{2} \).
Then \( u_{n}\in H^{1}(\Omega _{1}) \) , \( v_{n}\in H^{1}(\Omega _{2}) \),
\( \Vert u_{n}\Vert _{H^{1}(\Omega _{1})}\leq 1 \), \( \Vert v_{n}\Vert _{H^{1}(\Omega _{2})}\leq 1 \).

Because the embedding operator \( H^{1}(\Omega _{1})\to L^{2}(\Omega
_{1}) \)
is compact we can choose a subsequence \( \{u_{n_{k}}\} \) of the sequence
\( \{u_{n}\} \) which converges in \( L^{2}(\Omega _{1}) \) to a function
\( u_{0}\in L^{2}(\Omega _{1}) \). Because the second embedding operator \( H^{1}(\Omega _{2})\Rightarrow L^{2}(\Omega _{2}) \)
is also compact we can choose a subsequence \( \{v_{n_{k_{m}}}\} \) of the
sequence \( \{v_{n_{k}}\} \) which converges in \( L^{2}(\Omega _{2}) \) to
a function \( v_{0}\in L^{2}(\Omega _{2}) \). It is 
evident that \( u_{0}=v_{0} \)
almost everywhere in \( \Omega _{1}\cap \Omega _{2} \) and the function \( w_{0}(x) \)
which is defined as \( w_{0}(x):=u_{0}(x) \) on \( \Omega _{1} \) and \( w_{0}(x):=v_{0}(x) \)
on \( \Omega _{2} \) belongs to \( L^{2}(\Omega _{1}\cup \Omega _{2}) \).

Hence

\[
\Vert w_{n_{k_{m}}}-w_{0}\Vert _{L^{2}(\Omega _{1}\cup \Omega _{2})}\leq \Vert u_{n_{k_{m}}}-u_{0}\Vert _{L^{2}(\Omega _{1})}+\Vert v_{n_{k_{m}}}-v_{0}\Vert _{L^{2}(\Omega _{2})}.\]

Therefore \( \Vert w_{n_{k_{m}}}-w_{0}\Vert _{L^{2}(\Omega _{1}\cup \Omega _{2})}\rightarrow 0 \)
for \( m\rightarrow \infty  \) .
\end{proof}
From Proposition \ref{ecomp} and Lemma \ref{sob}  the main
result of this section  follows  immediately:

\begin{thm}
\label{2main}If a domain \emph{\( \Omega  \)} belongs to class \( T \) then
the embedding operator \( H^{1}(\Omega )\to L^{2}(\Omega ) \) is compact.
\end{thm}
The example below demonstrates the difference between class \( T \) and the
class of bounded domains whose boundaries are 
locally graphs of continuous functions ($C$-domains).
The boundary of a domain of class \( T \) can have countably many connected
components, while this is not possible for $C$-domains.

\begin{example} \label{component}
Take: \( U:=\{(x_{1},x_{2}):0<x_{1}<1/\pi ,x_{1}\sin \frac{1}{x_{1}}<x_{2}<x_{1}\sin \frac{1}{x_{1}}+4\} \);
\( V=(0,1/\pi )\times (-2,0) \) , \( \Omega =U\cup V \).

Domains \( U \) and \( V \) are elementary domains of class \( ET. \) Therefore
\( \Omega  \) is a domain of class \( T \) . By Theorem \ref{2main} the embedding
operator \( H^{1}(\Omega )\Rightarrow L^{2}(\Omega ) \) is compact.

Let us discuss the structure of \( \partial \Omega  \). The boundary \( \partial U \)
is connected and contains the graph
 \( \Gamma _{f}=\{(x_{1},x_{2}):x_{2}=x_{1}\sin \frac{1}{x_{1}} \)
of the function \( f:[0,\frac 1{\pi}]\rightarrow R \), \(
f(x_{1})=x_{1}\sin \frac{1}{x_{1}} \).
The graph \( \Gamma _{f} \) can be divided on two parts: the ``nonnegative''
part \( \Gamma _{f}^{+}:=\{(x_{1},x_{2})\subset \Gamma _{f}:x_{2}\geq 0\} \)
and ``negative part \( \Gamma _{f}^{-}:=\{(x_{1},x_{2})\subset \Gamma _{f}:x_{2}<0\} \).
The ``negative'' part \( \Gamma _{f}^{-}\subset V \). Therefore the boundary
\( \partial \Omega  \) of the plane domain \( \Omega  \) does not contain
\( \Gamma _{f}^{-} \) and consists of the countably many connected components:
\( S_{1}=([0,1/\pi ]\times \{-2\})\cup (\{0\}\times (-2,4))\cup (\{\frac{1}{\pi }\}\times (-2,4))\cup \Gamma _{g} \),
where \( \Gamma _{g} \) is the graph of the function
\( g:[0,\frac 1 {\pi}]\rightarrow R \)
, \( g(x_{1})=x_{1}\sin \frac{1}{x_{1}}+4 \); \( S_{i}=([\frac{1}{(2i-1)\pi },\frac{1}{2(i-1)\pi }]\times \{0\})\cup \Gamma _{i} \)
\( i=2,... \), \( \Gamma _{i}\subset \Gamma _{f}^{+} \) is the graph of the
restriction of the function \( f(x_{1})=x_{1}\sin \frac{1}{x_{1}} \) to the
segment \( [\frac{1}{(2i-1)\pi },\frac{1}{2(i-1)\pi }] \); and \( \widetilde{S}=\{0,0\} \)
is also a point of the boundary \( \partial \Omega  \).

Notice that any neighboorhood of the point \( \{0,0\} \) the boundary \( \partial \Omega  \)
has countably many connected components and therefore can not be presented as
a graph of any continuous function which is a connected set.

Higher-dimensional 
examples can be constructed using the rotation of  \\
two-dimensional
domain \( \Omega  \) around \( x_{1} \)-axis.
\end{example}
The following corollary is practically convenient for using the main theorem.

\begin{cor}
\label{union}If a bounded domain \( U \) is an extension domain, a domain
\( V \) belongs to class \( T \) and \( \Omega :=U\cup V \), then the embedding
operator \( H^{1}(\Omega )\to L^{2}(\Omega ) \) is compact.
\end{cor}
This corallary follows from Theorem \ref{2main} and Lemma \ref{sob}.

\begin{example}
Let \( U:=U(f,g,x^{'}_{0},r):=\{(x^{'},x_{n}):g(x^{'})<x_{n}<f(x^{'})\} \)
where a continuous real-valued functions \( f,g \) defined on the closed ball
\( \overline{B}:=\overline{B_{n-1}}(x^{'}_{0},r)\subset  \) \( R^{n-1} \)
and \( H:=\max _{x^{'}\in \overline{B}}(f(x^{'})-g(x^{'}))>0 \)\emph{.}.
Then
the embedding operator \( H^{1}(U)\to L^{2}(U) \) is compact.
\end{example}
The above claim follows from corollary \ref{union}. We need only to represent
\( U \) as a union of domains of class \( C \) and an extension domain (in
our case a domain with Lipschitz boundary).

\begin{rem*}
Extension domains can have very rough boundary. In the plane a bounded domain
\( U \) has an extension property if and only if it is an image of the unit
disc under quasiconformal homeomorphism \( \phi :R^{2}\rightarrow R^{2} \)
(see \cite{GV},\cite{GR}). For example the Hausdorff dimension of an image
\( \partial U \) of a unit circle under quasiconformal homeomorphism \( \phi :R^{2}\rightarrow R^{2} \)
can be any number \( 1\leq \alpha <2 \) \cite{GhV}.
\end{rem*}

\section{Quasiisometrical homeomorphisms and compact embeddings.}

A large class of bounded domains in \( R^{n} \) does not belong to class \( T \)
but still have ``good'' properties like compactness of the 
embedding \( H^{1}(\Omega )\Rightarrow L^{2}(\Omega ) \).
To study these domains we will introduce a larger and more flexible class of
``elementary'' domains, i.e. quasiisometrical images of elementary domains
of class \( ET. \) 
Then we extend the main theorem to the finite unions of quasiisometrical
elementary domains. Our proof 
is based on the well-known fact that a quasiisometrical
homeomorphism \( \varphi :U\rightarrow V \) induces 
a bounded composition operator
\( \varphi ^{*}:H^{1}(V)\to H^{1}(U) \) by the rule \( \varphi
^{*}(u)=u\circ \varphi  \)
(see, for example \cite{GR} or \cite{Z}).

Recall the definition of a quasiisometrical homeomorphism.

\begin{defn}
Let \( U \) and \( V \) be two domains in \( R^{n} \). A
homeomorphism \( \varphi :U\rightarrow V \)
is \( Q- \)quasiisometrical (or simply quasiisometrical) if for any point \( x\in U \)
there exists such a ball \( B(x,r)\subset U \) that
\begin{equation}
\label{3}
Q^{-1}|y-z|<|\varphi (y)-\varphi (z)|<Q|y-z|
\end{equation}
 for any \( y,z\in B(x,r) \). Here the constant \( Q>0 \) does not depend
on the choice of \( x\in U \).
\end{defn}
Obviously the inverse homeomorphism \( \varphi ^{-1}:V\rightarrow U \) is also
\( Q- \)quasiisometrical . Domains \( U \) and \( V \) are quasiisometrically
equivalent if there exists a quasiisometrical homeomorphism \( \varphi :U\rightarrow V \).

Any quasiisometrical homeomorphism is a locally bi-Lipschitz, weakly differentiable
and differentiable almost everywhere.

Any diffeomorphism \( \varphi :U\rightarrow V \) is quasiisometrical on a subdomain
\( U_{1}\subset U \) if the closure \( \overline{U_{1}} \) of \( U_{1} \)
belongs to \( U \) .

Let us demonstrate a practical way to construct a new quasiisometrical homeomorphism
using a given one. Suppose that \( S_{k}(x)=kx \) is a similarity transformation
(which is called below a similarity) of \( R^{n} \) with the similarity coefficient
\( k>0 \), \( S_{k_{1}}(x)=k_{1}x \) is
 another similarity and \( \varphi :U\rightarrow V \)
is a \( Q- \)quasiisometrical homeomorphism. Then a composition \( \psi :=S_{k}\circ \varphi \circ S_{k_{1}} \)
is a \( k_{1}kQ \) -quasiisometrical homeomorphism.

It is easy to check this claim. Because \( \varphi :U\rightarrow V \) is \( Q- \)quasiisometrical
for any point \( x\in U \) there exists such a ball \( B(x,r)\subset U \)
that the inequality \ref{1} holds.

Therefore
\[
|\psi (y)-\psi (z)|=k|\varphi (k_{1}y)-\varphi (k_{1}z)|<kQ|k_{1}y-k_{1}z|<k_{1}kQ|y-z|\]
 for any \( y,z\in k_{1}^{-1}B(k_{1}^{-1}x,k_{1}^{-1}r) \). By the same reasons
\[
|\psi (y)-\psi (z)|>(k_{1}kQ)^{-1}|y-z|.\]

If \( k_{1}=k^{-1} \) then the homeomorphism \( \psi  \) is \( Q- \)quasiisometrical.

This remark will be used in example 2.2 of a domain with ``spiral'' boundary
which is quasiisometrically equivalent to a cube. We start with a two-dimensional
Wexample.

\begin{example}
We will construct a domain with ``spiral'' boundary with the help of a quasiisometrical
homeomorphism. We can start with the triangle \( T:=\{(s,t):0<s<1,s<t<2s\} \)
because \( T \) is quasiisometrically equivalent to the unit square \( Q_{2}=(0,1)\times (0,1) \).
Hence we need to construct only a quasiisometrical homeomorphism \( \varphi _{0} \)
from \( T \) into \( R^{2} \).

Let \( (\rho ,\theta ) \) be polar coordinates in the plane. Define first
a mapping \( \varphi :R_{+}^{2}\rightarrow R^{2} \) as follows: \( \varphi (s,t)=(\rho (s,t),\theta (s,t)) \),
\( \rho (s,t)=s \) , \( \theta (s,t)=2\pi \ln \frac{t}{s^{2}} \). Here \( R_{+}^{2}:=\{(s,t)|0<s<\infty ,0<t<\infty \} \).
An inverse mapping can be calculated easily: \( \varphi ^{-1}(\rho ,\theta )=(s(\rho ,\theta ),t(\rho ,\theta )) \),
\( s(\rho ,\theta )=\rho  \), \( t(\rho ,\theta ))=\rho ^{2}e^{\frac{\theta }{2\pi }} \).
Therefore \( \varphi  \) and \( \varphi _{0}=\varphi |T \) are diffeomorphisms.

The image of the ray \( t=ks, \) \( s>0,k>0 \) is the
logarithmic spiral \( \rho =k\exp (-\frac{\theta }{2\pi }) \). Hence the image \( S:=\varphi (T)=\varphi _{0}(T) \)
is an ``elementary spiral'' plane domain, because \( \partial T \) is a union
of two logarithmic spirals \( \rho =\exp (-\frac{\theta }{2\pi }) \), \( \rho =2\exp (-\frac{\theta }{2\pi }) \)
and the segment of the circle $\rho=1$ .

The domain \( T \) is a union of countably many subdomains \( T_{n}:=\{(s,t):e^{-(n+1)}<s<e^{-(n-1)},s<t<2s\} \),
\( n=1,2,... \) . On the first domain \( T_{1} \) the diffeomorphism \( \varphi _{1}:=\varphi |T_{1} \)
is \( Q- \)quasiisometrical, because \( \varphi _{1} \) is the restriction
on \( T_{1} \) of a diffeomorphism \( \varphi  \) defined in \( R_{+}^{2} \)
and \( \overline{T_{1}}\subset R_{+}^{2} \). We do not calculate the number
\( Q \).

If we will prove that any diffeomorphism \( \varphi _{n}:=\varphi |T_{n} \)
is the composition \( \varphi _{n}=S_{e^{-(n-1)}}\circ \varphi _{1}\circ S_{e^{n-1}} \)
of similarities \( S_{e^{-(n-1)}} \), \( S_{e^{n-1}} \) and the \( Q-
\)quasiisometrical
diffeomorphism \( \varphi _{1} \), then any diffeomorphism \( \varphi _{n} \)
is \( Q- \)quasiisometrical, the diffeomorphism \( \varphi _{0} \) is also
\( Q- \)quasiisometrical, and the ``elementary spiral'' domain \(
U=\varphi _{0}(T) \)
is quasiisometrically equivalent to the unit square.

Let us prove the representation \( \varphi _{n}=S_{e^{-(n-1)}}\circ \varphi _{1}\circ S_{e^{n-1}} \).

By construction the domain \( T_{1} \) is the image of \( T_{n} \) under the
similarity transformation \( S_{e^{n-1}}(s,t)=e^{n-1}(s,t) \). Therefore we
need to prove only the representation \( \varphi =S_{e^{-(n-1)}}\circ \varphi \circ S_{e^{n-1}} \).
This representation follows from a direct calculation:
\[
(S_{e^{-(n-1)}}\circ \varphi \circ S_{e^{n-1}})(s,t)=S_{e^{-(n-1)}}(\rho (e^{n-1}s,e^{n-1}t),\theta (e^{n-1}s,e^{n-1}t))\]
\[
=(e^{-(n-1)}\rho (e^{n-1}s,e^{n-1}t),\theta (e^{n-1}s,e^{n-1}t))=(s,2\pi \ln \frac{t}{s^{2}}-2\pi (n-1))\]
\[
=(\rho (s,t),\theta (s,t))=\varphi (s,t)\]

\end{example}
\begin{rem*}
By a rotation we can construct corresponding higher-dimensional examples
of domains
with ``spiral'' type singularities.
\end{rem*}

\subsection{Domains of class \protect\protect\protect\protect\protect\( L\protect \protect \protect \protect \protect \).}

\begin{defn}
\emph{A domain U is an elementary domain of class \( L \) if it is a quasiisometrical
image of an elementary domain of class \( ET \).}

\emph{A domain U is a domain of class \( L \) if it is a
finite union of elementary
domains of class \( L \).}
\end{defn}
\begin{prop}
(see for example \cite{GR} or \cite{Z}) Let \( U \) and \( V \)
be domains in \( R^{n} \).
A quasiisometrical homeomorphism \( \varphi :U\rightarrow V \) induces a bounded
composition operator \( \varphi ^{*}:H^{1}(V)\Rightarrow H^{1}(U) \) by the
rule \( \varphi ^{*}(u)=u\circ \varphi  \).
\end{prop}
Combining this result with Theorem \ref{2main} and Lemma \ref{sob} we
obtain:

\begin{thm}
\label{lmain}If a domain \emph{\( \Omega  \)} belongs to class \( L \) then
the embedding operator \( H^{1}(\Omega )\to L^{2}(\Omega ) \) is compact.
\end{thm}
\begin{proof}
Let \( U \) be an elementary domain of class \( L \). Then there exists an
elementary domain \( V \) of class \( ET\) and a quasiisometrical homeomorphism
\( \varphi :V\rightarrow U \). By the previous theorem operators \( \varphi ^{*}:H^{1}(V)\Rightarrow H^{1}(U) \)
and \( (\varphi ^{-1})^{*}:H^{1}(U)\to H^{1}(V) \) are bounded. By
Theorem \ref{ecomp} the embedding operator \( I_{V}:H^{1}(V)\to 
L^{2}(V) \)
is compact. The embedding operator \( I_{U}:H^{1}(U)\to L^{2}(U) \)
is the composition \( (\varphi ^{-1})^{*}\circ I_{V}\circ \varphi ^{*} \).
Therefore the embedding operator \( I_{U}:H^{1}(U)\to L^{2}(U) \) is
compact.

Because any domain \emph{\( \Omega  \)} of class \( L \) is a finite union
of elementary domains of class \( L \) the result follows from Lemma \ref{sob}.
\end{proof}

\section{Domains with nonlocal singularities of the boundaries}

The previous section focuses on domains which are locally quasiisometrical
images
of domains of class \( T. \) For the proof of the 
main result we used the compactness
of embedding operators for domains of class \( T \) and the boundedness of
composition operators induced by quasiisometrical homeomorphisms.

In this section we use similar arguments for the largest class of homeomorphisms
that induce bounded composition operators of the Sobolev spaces \( H^{1}
\).

We recall the main idea for a study of the embedding operators proposed
in \cite{GG}.
Let \( \Omega  \) be a domain with ``good'' boundary, for example, domain
of class \( L \), and \( U \) be a domain with ``bad'' boundary. Suppose
that there exists a homeomorphism \( \phi :\Omega \rightarrow U \) such that
\( \phi  \) induces a bounded composition operator \( \phi ^{*}:H^{1}(U)\rightarrow H^{1}(\Omega ) \)
by the rule \( \phi ^{*}(u)=u\circ \varphi  \) and \( \phi ^{-1} \) induces
a bounded composition operator \( (\phi ^{-1}) ^{*}:L^{2}(\Omega )\rightarrow L^{2}(U) \).
If the embedding operator \( I_{\Omega }:H^{1}(\Omega )\to L^{2}(\Omega
) \)
is compact, then the embedding operator \( I_{U}=(\phi ^{-1})^{*}I_{\Omega }\phi ^{*}:H^{1}(\Omega )\Rightarrow L^{2}(\Omega ) \)
is also compact.

This method was used in \cite{GG} for a study of the embedding operators
in domains with
``nonlocal'' singularities.

\subsection{2-quasi-conformal homeomorphisms.}

 Composition operators for Sobolev spaces with first generalized derivatives
were studied in detail in \cite{GGR}. We restrict
 ourselves to the practically
important class of locally bi-Lipschitz homeomorphisms.

\begin{defn}
\emph{A locally bi-Lipschitz homeomorphism} \( \phi :\Omega \rightarrow U \)
\emph{is 2-quasi-conformal if there exists a constant \( K \) such that
\[
\Vert \phi '(x)\Vert ^{2}\leq K|\det \phi '(x)|\]
 for almost all \( x\in \Omega  \) . The 2-quasi-conformal dilatation \( K(\phi ) \)
is a minimal number \( K \) for which the previous inequality holds.}
\end{defn}
Here \( \phi '(x)=|\frac{\partial \varphi _{i}}{\partial x _{j}}(x)|,i,j=1,2,..,n \)
is the Jacobi matrix of the mapping \( \varphi  \) at the point \( x \) and
\( \Vert \phi '(x)\Vert :=\sqrt{\sum _{i,j=1}^{n}(\frac{\partial \varphi _{i}}{\partial x _{j}}(x))^{2}} \)
is the norm of the Jacobi matrix.

Obviously any quasiisometrical homeomorphism is 2-quasi-conformal. Composition
of 2-quasi-conformal homeomorphisms is 2-quasi-conformal \cite{GG}.

Choose two bounded domains \( \Omega ,U \) in \( R^{n} \) , \( n>2 \) .

\begin{thm}
(see \cite{GG})\label{qcomp}A locally bi-Lipschitz homeomorphism \( \phi :\Omega \rightarrow U \)
induces a bounded composition operator \( \phi ^{*}:H^{1}(U)\rightarrow H^{1}(\Omega ) \)
if and only if \( \phi  \) is 2-quasi-conformal.
\end{thm}
This result was used in the following version of the so-called
``relative''
embedding theorem.

\begin{thm}
(see \cite{GGR}) Suppose that a homeomorphism \( \phi :\Omega \rightarrow U \)
is 2-quasi-conformal and \( \Vert \det \phi '(x)\Vert _{L^{\infty }(\Omega )}<\infty  \)
. If the embedding operator \( I_{\Omega }:H^{1}(\Omega )\to L^{2}(\Omega
) \)
is compact then the embedding operator \( I_{U}:H^{1}(U)\to L^{2}(U) \)
is also compact.
\end{thm}
The following corollary helps to use this result practically:

\begin{cor}
\label{eqcom} Suppose that \( \Omega  \) is domain of class \( L \) and there
exists a 2-quasi-conformal homeomorphism \( \phi :\Omega \rightarrow U \).
If \( \Vert \det \phi '(x)\Vert _{L^{\infty }(\Omega )}<\infty  \), then the
embedding operator \( I_{U}:H^{1}(U)\to L^{2}(U) \) is compact.
\end{cor}
This corollary follows immediately from the previous theorem and the embedding
theorem for \( T- \)domains.

It allows one to use the method of Section 2 for 2-quasi-conformal case.

\subsection{Domains of class \protect\protect\protect\protect\protect\( Q\protect \protect \protect \protect \protect \).}

\begin{defn}
\emph{A domain \( U \) is an elementary domain of class \( Q \) if there exist
an elementary domain \( V \) of class \( L \) and a 2-quasi-conformal homeomorphism
\( \varphi :U\rightarrow V \) such that} \( \Vert \det \phi '(x)\Vert _{L^{\infty }(\Omega )}<\infty  \)\emph{. }

\emph{A domain \( U \) is a domain of class \( Q \) if it is a finite union
of elementary domains of class \( Q \).}
\end{defn}
Combining Corollary \ref{eqcom} with the Theorem \ref{lmain} and Lemma \ref{sob}
we obtain

\begin{thm}
\label{qmain}If a domain \emph{\( \Omega  \)} belongs to class \( Q \) then
the embedding operator \( H^{1}(\Omega )\Rightarrow L^{2}(\Omega ) \) is compact.
\end{thm}
\begin{proof}
Let \( U \) be an elementary domain of class \( Q \). Then there exists an
elementary domain \( V \) of class \( L \) and a 2-quasiisometrical homeomorphism
\( \varphi :V\rightarrow U \) such that \( \Vert \det \phi '(x)\Vert _{L^{\infty }(\Omega )}<\infty  \)\emph{.}
By Theorem \ref{qcomp} operators \( \varphi ^{*}:H^{1}(V)\to H^{1}(U) \)
and \( (\varphi ^{-1})^{*}:H^{1}(U)\Rightarrow H^{1}(V) \) are bounded. By
Corollary \ref{eqcom} the embedding operator \( I_{V}:H^{1}(V)\to 
L^{2}(V) \)
is compact. The embedding operator \( I_{U}:H^{1}(U)\to L^{2}(U) \)
is equal to the composition \( (\varphi ^{-1})^{*}\circ I_{V}\circ \varphi ^{*} \).
Therefore the embedding operator \( I_{U}:H^{1}(U)\to L^{2}(U) \) is
compact.

Because any domain \emph{\( \Omega  \)} of class \( Q \) is a finite union
of elementary domains of class \( Q \), the result follows from Lemma \ref{sob}.
\end{proof}
Let us demonstrate a simple example of an elementary domain of class \( Q \)
with ``non local'' singularity near the point \( \{0\} \).

\begin{example}
Let \( \Omega \in R^{2} \) be the union of rectangles \( T_{k}=\{x\in R^{2}:|x_{1}-2^{-\alpha k}|\leq 2^{-\alpha (k+2)};0\leq x_{2}<2^{-\alpha (k+2)}\},0<\alpha  \)
and the square \( Q=(0,1)\times (-1,0) \) . It is easy to check that the homeomorphism
\( \varphi (x_{1},x_{2})=(x_{1}|x|^{\frac{1}{\alpha }-1},x_{2}|x|^{\frac{1}{\alpha }-1}\} \)
is 2-quasi-conformal and \( \Omega _{1}=\varphi (\Omega ) \) is the union of
rectangles \( P_{k}=\{x\in R^{2}:|x_{1}-2^{-k}|\leq 2^{-(k+2)};0\leq x_{2}<2^{-(k+2)}\},0<\alpha  \)
and the square \( Q=(0,1)\times (-1,0) \) . In \cite{M} a quasiisometrical
homeomorphism \( \psi  \) from \( \Omega _{1} \) to the unit square is constructed.
Hence the composition \( \phi =\psi \circ \varphi  \) is a 2-quasi-conformal
homeomorphism and by direct calculation we can check that \( \Vert \det \phi '(x)\Vert _{L^{\infty }(\Omega )}<\infty  \)\emph{.}
Therefore the domain \( \Omega  \) is an elementary domain of class \( Q \).

Remark that a projection of \( B(0,r)\cap \partial \Omega  \) into arbitrary
line \( L\in R^{2} \) is not one to one correspondence for any \( r \) and
\( L \) . Therefore the domain \( \Omega  \) is not an elementary domain of
class \( C \).

Higher-dimensional examples can be constructed using rotations.
\end{example}

\subsection{Discussion of 2-quasiconformal homeomorphisms and \newline 2-quasi-conformal
domains.}

Let us give first a geometrical interpretation of 2-quasi-conformality.

Suppose that \( \phi :R^{n}\rightarrow R^{n} \) is a linear homeomorphism,
\( \varphi ' \) is its matrix and \( (\phi ')^{T} \) its adjoint matrix. Denote
by \( \lambda _{1}^{2}\leq \lambda _{2}^{2}\leq ...\leq \lambda _{n}^{2} \)
eigenvalues of \( (\phi ')^{T}\phi ' \). There exist two orthogonal bases \( e_{1},e_{2},...,e_{n} \)
and \( g_{1},g_{2},...,g_{n} \) such that \( \phi (e_{i})=\lambda _{i}g_{i} \)
for every \( i=1,2,...,n \). Geometrically \( \lambda _{i} \) is length of
\( i- \)th semi-axis of the ellipsoid \( \phi (B(0,1)) \) . The
2-quasi-conformal
dilatation \( K(\phi )=\frac{\lambda _{n}}{\lambda _{1}\lambda _{2}...\lambda _{n-1}} \).

If \( \varphi :\Omega \rightarrow U \) is a diffeomorphism then the numbers
\( \lambda _{1}(x)\leq \lambda _{2}(x)\leq ...\leq \lambda _{n}(x) \) correspond
to the linear homeomorphism \( d\phi  \) and
\[
K(\phi )=\sup _{x\in \Omega }[\frac{\lambda _{n}(x)}{\lambda _{1}(x)\lambda _{2}(x)...\lambda _{n-1}(x)}].\]

If \( \varphi :\Omega \rightarrow U \) is only locally Lipschitz then
\[
K(\phi )=esssup_{x\in \Omega }[\frac{\lambda _{n}(x)}{\lambda _{1}(x)\lambda _{2}(x)...\lambda _{n-1}(x)}].\]

The relations of 2-quasi-conformal homeomorphisms with the traditional classes
can be described as follows:

1) In the two-dimensional case 2-quasi-conformal homeomorphisms are
quasi-conformal.
A homeomorphism inverse to a quasi-conformal homeomorphism is also quasi-conformal.
Therefore a homeomorphism inverse to 2-quasi-conformal homeomorphism is 2-quasi-conformal
(for plane domains). Unfortunately, this property does not hold in the higher-dimensional
cases. In \cite{ GGR} an example of 2-quasi-conformal homeomorphism with non-2-quasi-conformal
inverse homeomorphism is constructed. Composition of 2-quasi-conformal homeomorphisms
is a 2-quasi-conformal homeomorphism.

2) Two-dimensional conformal mappings are 2-quasi-conformal homeomorphisms with
\( K(\phi )=1 \).

3) Any quasiisometrical homeomorphism is 2-quasi-conformal.

\section*{Appendix}

In this section an abstract necessary and sufficient condition for
the embedding operator to be compact is given.
In our presentation the work \cite{Ra} is used.
 
Let \( H_{j}, \) \( j=1,2,3, \) be Hilbert spaces,
 \( H_{1}\subset H_{2}\subset H_{3}, \)
 the embeddings mean set-theoretical inclusions 
and the inequalities \( ||u||_{1}\geq ||u||_{2}\geq ||u||_{3} \),
where \( ||u||_{j}:=||u||_{H_{j}} \). This implies the compatibility of the
norms: 

if \( ||u_{n}||_{3}\rightarrow 0 \) and \( ||u_{n}-u||_{2}\rightarrow 0 \)
then \( u=0 \) 

Denote by \( i \) the embedding operator from \( H_{1} \) into \( H_{2} \)
and by \( j \) the embedding operator from \( H_{1} \) into \( H_{3} \). 

\begin{prop}
\noindent The operator \( i: \) \( H_{1}\rightarrow H_{2} \) is compact if
and only if the following conditions hold: 

1) \( j \) is compact, 

and 

2) \( ||u||_{2}\leq \varepsilon ||u||_{1}+c(\varepsilon )||u||_{3} \) for all
\( \varepsilon \in (0,\varepsilon _{0}) \), \( c(\varepsilon )=const>0 \),
for all \( u\in H_{1} \) .
\end{prop}
\begin{proof}
\noindent \textit{\underbar{Necessity:}} condition 1) is clearly necessary:
if \( i \): \( H_{1}\rightarrow H_{2} \) is compact, and \( H_{2}\subset H_{3} \),
\( ||u||_{2}\geq ||u||_{3} \), then \( j \): \( H_{1}\rightarrow H_{3} \)
is compact.

To prove 2), assume the contrary: there exists \( u_{n}\in H_{1} \), \( ||u_{n}||_{1}=1 \),
and \( \varepsilon \in (0,\varepsilon _{0}) \) such that
\[
||u||_{2}>\varepsilon ||u||_{1}+n||u||_{3}\]
 for all \( n=1,2,... \).

Since \( ||u_{n}||_{1}=1\geq ||u_{n}||_{2} \), one concludes from previous
inequality that \( ||u_{n}||_{3}\rightarrow 0 \) as \( n\rightarrow \infty  \)
and \( u_{n}\rightarrow u \) in \( H_{1} \), \( \rightarrow  \) stands for
weak convergence. Since \( i \): \( H_{1}\rightarrow H_{2} \) is compact,
it follows that \( ||u_{n}-u||_{2}\rightarrow 0. \) Since \( ||u_{n}||_{3}\rightarrow 0 \)
it follows that \( u=0 \) and \( ||u_{n}||_{2}\rightarrow 0 \). This is a
contradiction: by condition 2) the inequality \( ||u_{n}||_{2}\geq \varepsilon >0 \)
holds. Necessity of 1) and 2) is established.

\noindent \textit{\underbar{Sufficiency:}} if \( j \) is compact then \( ||u_{n}||_{1}=1 \)
implies that a subsequence \( u_{n} \) (denoted again \( u_{n} \)) converges
in \( H_{3} \), that is \( ||u_{n}-u_{m}||_{3}\rightarrow 0 \) as \( n,m\rightarrow \infty  \).
Condition 2) implies \( ||u_{n}-u_{m}||_{2}\leq \varepsilon ||u_{n}-u_{m}||_{1}+c(\varepsilon )||u_{n}-u_{m}||_{3} \).

Fix an arbitrary small \( \delta >0 \). Note that \( ||u_{n}-u_{m}||_{1}\leq 2 \).
Choose \( \varepsilon =\delta /4 \) and fix it. Then choose \( n,m \) so large
that \( c(\varepsilon )||u_{n}-u_{m}||_{3}<\delta /2 \). Then \( ||u_{n}-u_{m}||_{2}<\delta  \).
This implies convergence of \( u_{n} \) in \( H_{2} \). 
The sufficiency is proved.
\end{proof}

Acknowledgement: AGR thanks Prof. V.Maz'ya for useful correspondence.

\end{document}